\documentstyle[preprint,aps]{revtex} 

\begin{document}
\draft

\title{Modeling of Financial Data: Comparison of the Truncated L\'evy Flight and the ARCH(1) and GARCH(1,1) processes}

\author{Rosario N. Mantegna$^1$ and H. Eugene Stanley$^2$}

\address{$^1$Istituto Nazionale per la Fisica della Materia, Unit\`a di
Palermo and Dipartimento di Energetica ed Applicazioni di Fisica,
Universit\`a di Palermo, Palermo, I-90128, ITALIA\\ $^2$Center for
Polymer Studies and Dept. of Physics, Boston University, Boston, MA
02215 USA}

%\date{\today}

\maketitle

\begin{abstract}
 We compare our results on empirical analysis of 
financial data with simulations of two stochastic models of the dynamics
of stock market prices. The two models are (i) the truncated L\'evy flight recently introduced by us and (ii) the ARCH(1) and GARCH(1,1) 
processes. 
We find that the TLF well
describes the scaling and its breakdown observed in empirical data, 
while it is not able to properly describe the fluctuations of volatility empirically
detected. The ARCH(1) and GARCH(1,1) models are able to describe the 
probability density function of price changes at a given time horizon,
but both fail to describe the scaling properties of the PDFs for
short time horizons. 
\end{abstract}

\newpage

\section{Introduction}

Inspired by pioneering works on the analysis and modeling of economic
and financial systems \cite{Mandelbrot63,Kadanoff71,Montroll74}, a growing number of physicists are becoming involved in the analysis and modeling of financial markets 
\cite{Mantegna91,Li91,Takayasu92,Bak93,Bouchaud94,Mantegna95,Ghashghaie96,Mantegna96,ArneodoP,Solomon96,GalluccioPA,BakPA,Potters,VandewallePA,Liu97,Cizeau97}.
In this lecture we consider the price
dynamics of a stock index traded in a financial market. The most accepted paradigm in finance is that no arbitrage
is present in financial markets i.e. there is no way to extract 
money from the market continuously and without risk \cite{Ingersoll87}.

In this lecture, we firstly recall results obtained by us 
\cite{Mantegna95,Mantegna96}, by performing an empirical analysis 
of high-frequency data of one of the most important indices of the
New York Stock Exchange, the Standard \& Poor's 500 (S\&P500) index.
The results obtained in the empirical analysis are used as benchmarks
for two stochastic processes used to model price dynamics in financial
markets. The first model is the truncated L\'evy flight (TLF), recently
introduced by us \cite{Mantegna94}. The second stochastic process
belongs to the class of autoregressive conditional heteroscedasticity
(ARCH) models \cite{Engle82} and to its generalization (GARCH)
\cite{Bollerslev86}. We address strengths and weaknesses of all three 
models in describing real financial data. 
We focus our attention on the probability density functions (PDFs)
of price changes at different time horizons, on the scaling 
properties of the PDFs and on the degree of stationarity of index 
changes. 
\section{Empirical analysis}

We performed empirical analyses of the dynamics of indices of 
stock prices traded in financial markets \cite{Mantegna91,Mantegna95,Mantegna96}.
Our
empirical analysis \cite{Mantegna95,Mantegna96} of the S\&P500 Index
of the New York Stock Exchange shows that a non-Gaussian scaling of the 
PDF of price changes is present at short 
times (from $\Delta t=1$ to $\Delta t=1000$ trading minutes) while
a breakdown from the non-Gaussian scaling is present for long times 
($\Delta t >> 1000$ trading minutes) \cite{Mantegna95,Mantegna96}.
We performed our analysis by analyzing high frequency data recorded
during the 6-year period 1/84--12/89 (time intervals between
successive records as short as 15~seconds are present in the data base). 
In our analysis \cite{Mantegna95}, we define the trading time as a continuous time starting
from the opening of the day until the closing, and then continuing with
the opening of the next trading day. From this data base, we select the
complete set of non-overlapping records separated by a time interval
$\Delta t\pm\epsilon\Delta t$ (where $\epsilon$ is the tolerance, always
less than 0.035).  We denote the value of the S \& P 500 as $y(t)$, and
we define $z(t)\equiv y(t)-y(t-\Delta t)$. In our analysis, we
determine \cite{Mantegna95} the probability distribution $P(z)$ of index
variations for different values of $\Delta t$. We select $\Delta t$
values that are logarithmically equally spaced ranging from 1 to 1000
min. The number of data in each set is decreasing from the maximum
value of 493,545 ($\Delta t=1$ min) to the minimum value of 562 ($\Delta
t=1000$ min). We note \cite{Mantegna95} that the distributions are non-Gaussian, indeed, they have wings larger than expected for a normal process. 

We study the ``probability of
return to the origin'' $P(z=0)$ as a function of $\Delta t$. With this
choice, we are investigating the point of each probability distribution that is
{\it least\/} affected by the noise introduced by the finiteness of the
experimental data set. Our investigation of $P(0)$ versus $\Delta t$ in
a log-log plot \cite{Mantegna95} shows that the data are well-fit by a
straight line characterized by the slope $-0.712\pm 0.025$. We observe
a non-normal scaling behavior (slope $\neq-0.5$) in an interval of
trading time ranging from 1 to 1000 min.

For short time horizons (from $\Delta t=1$ to $\Delta t=1000$ minutes), this empirical finding agrees with the model of  a L\'evy
flight proposed by Mandelbrot in 1963 to model cotton price dynamics \cite{Mandelbrot63} or with the model of a L\'evy
walk \cite{Shlesinger93}. In fact, if the
central region of the distribution is well-described by a L\'evy stable
symmetrical distribution \cite{Levy37},
\begin{equation}
L_\alpha(z,\Delta t)\equiv{1\over\pi}\int_0^\infty\exp(-\gamma\Delta
tq^\alpha)\cos(qz)dq,
\end{equation}
of index $\alpha$ and scale factor $\gamma$ at $\Delta t=1$, then the
probability of return is given by
\begin{equation}
P(0)\equiv L_\alpha(0,\Delta
t)={\Gamma(1/\alpha)\over\pi\alpha(\gamma\Delta t)^{1/\alpha}}.
\end{equation}
By using the value $-0.712$ from the analysis of the probability of
return, we obtain the index $\alpha=1.40\pm 0.05$ \cite{Mantegna95}. 

We also check if the scaling extends over the entire probability
distribution as well as $z = 0$. 
All the distributions (with $\Delta t=1$ to $\Delta t=1000$ minutes) agree well with a L\'evy stable distribution
\cite{Mantegna95,Mantegna97}. The
distributions obtained with the highest temporal resolution ($\Delta t <
10$) show that in addition to the good agreement with the L\'evy
(non-Gaussian) profile observed for almost three orders of magnitude, an
approximately exponential fall-off is present. The clear deviation of
the tails of the distribution from the L\'evy profile shows that the
experimental tails are less fat than expected for a L\'evy
distribution. 

The L\'evy distribution has an infinite second moment (if $\alpha < 2$)
\cite{Levy37}. However, our empirical finding of an exponential (or
stretched exponential) fall-off implies that the second moment is
finite. This
conclusion might at first sight seem to contradict our observation of
L\'evy scaling of the central part of the price difference distribution
over fully three orders of magnitude. However, the contradiction
is more apparent than real since, for example, the above findings are consistent with the 
theoretical predictions of the truncated L\'evy flight \cite{Mantegna94}.
 
\section{The Truncated L\'evy Flight}

The truncated L\'evy flight (TLF) has been introduced by Mantegna and
Stanley in Ref.~\cite{Mantegna94}. A TLF is defined as a stochastic
process $\{x\}$ characterized by the following probability density
function
\begin{equation}
\label{e.2}
T(x)\equiv\cases{
0      & $x > \ell$  \cr
   c_1L(x) & $-\ell\leq x\leq \ell$ \cr
0      & $x < -\ell$  \cr },
\end{equation}
where $L(x)$ 
is the symmetrical L\'evy stable distribution of index $\alpha$ $(0 <
\alpha\leq 2)$ and scale factor $\gamma$ $(\gamma > 0)$, $c_1$ is a
normalizing constant and $\ell$ is the cutoff length. In the following
theoretical considerations, for the sake of
simplicity, we set $\gamma=1$.

The central limit theorem (CLT) is fundamental to statistical
mechanics. It states that when $n\to\infty$, the sum
\begin{equation}
\label{e.1}
z_n\equiv\sum_{i=1}^nx_i
\end{equation}
of $n$ stochastic variables $\{x\}$ that are statistically independent,
identically distributed and with a finite variance, converges to a
normal (Gaussian) stochastic process. Generally, $n\approx10$ is
sufficient to ensure convergence. In a dynamical system, Eq.~(\ref{e.1})
defines a random walk if the variable $x$ is the jump size performed
after a time interval $\Delta t$ and $n$ is the number of time
intervals. Here, the ``number of variables'' $n$ and the
``time'' $t=n \Delta t$ can be interchanged everywhere.

For low values of $n$, $P(z_n=0)$ takes a value very close to the one
expected for a L\'evy stable process
\begin{equation}
\label{e.4}
P(z_n=0)\simeq L(z_n=0)={\Gamma(1/\alpha)\over\pi
\alpha n^{1/\alpha}}.
\end{equation}
For large values of $n$, $P(z_n=0)$ assumes the value predicted for a
normal process,
\begin{equation}
\label{e.5}
P(z_n=0)\simeq N(z_n=0)={1\over\sqrt{2\pi }\sigma_o(\alpha,\ell)n^{1/2}},
\end{equation}
where $\sigma_o(\alpha,\ell)$ is the standard deviation of the TLF
stochastic process $\{x\}$.

In the interval $1\leq\alpha<2$, the crossover between the two regimes
has been determined in Ref. \cite{Mantegna94} as:
\begin{equation}
n_{\times}\approx A \: \ell^\alpha,
\end{equation}
where $A$ is a function of $\alpha$ (the explicit form is given in 
Ref. \cite{Mantegna94}).
The description of the convergence process does not depend crucially on
the exact shape of the cut-off \cite{Shlesinger95} and some results of
Ref.~\cite{Mantegna94} have been confirmed analytically for an
exponential cut-off in Ref.~\cite{Koponen95}.

By performing numerical simulations, we verified \cite{Mantegna94} that 
the probability of return to the origin indicates
with high accuracy the degree of convergence of the process to one of
the two asymptotic regimes. 
  
The TLF model explains the empirical observations of (i) non-Gaussian 
scaling of the PDFs of price changes for short times; (ii) L\'evy 
shape of the {\it central part} of the price change distributions for $\Delta t \le 1000$ trading minutes; (iii) gradual convergence to a Gaussian 
process for long time horizons ($\Delta t >> 1000$ trading minutes).
However, not all the features observed in the S\&P 500 dynamics
are described by the TLF model. The simplest version of the model
cannot describe the short time memory (of the order of 20 minutes or
less) observed in the empirical data \cite{Mantegna96,Mantegna97} and also does
not explain the empirical observation of the time dependence of
the parameter $\gamma$ which is fluctuating with burst of activity
localized in specific months \cite{Mantegna95,Mantegna97}. The $\gamma$
parameter is related to what is called ``volatility" in the
economic literature \cite{Schwert89}.

\section{ARCH process}

ARCH stochastic models were introduced by Engle in 1982 \cite{Engle82}.
They are stochastic models with autoregressive conditional heteroscedasticity,
namely zero mean, uncorrelated stochastic processes with nonconstant variances
conditional on the past. These models have a very interesting property:
they might be locally unstationary (for short time intervals) but
globally stationary for well defined ranges of the values of the
control parameters. They are widely known in the economic literature
\cite{Bollerslev92}, but they are almost unknown to the physics community
in spite of the fact that they might also be useful in the description
of physical problems.

The simplest ARCH model is the ARCH(1) model defined as a random
variable $Z$ which is characterized at time $t$ by a variance
$\sigma_t^2$ given by
\begin{equation}
\sigma^2_t=\alpha_0+\alpha_1 Z_{t-1}^2
\end{equation}
where $Z_{t-1}$ is a random variable selected from a set of random 
variables characterized by a Gaussian distribution with zero mean
and standar deviation $\sigma_{t-1}$. $\alpha_0$ and $\alpha_1$ are the
control parameters of the stochastic process.

The most general ARCH stochastic process, the ARCH(n) process is 
defined by
\begin{equation}
\sigma^2_t=\alpha_0+\alpha_1 Z_{t-1}^2+ .............. + \alpha_n Z_{t-n}^2
\end{equation}
where $\alpha_0, .... , \alpha_n$ are control parameters and $Z_{t-1}, .... , Z_{t-n}$ are random variables drawn from sets of random variables
with Gaussian distributions of zero mean and standard deviations
$\sigma_{t-1}, .... , \sigma_{t-n}$ respectively. In spite of the fact that
$\sigma_t$ is showing an intermittent-like behavior, the overall process
$\{ Z\}$ is stationary on a long time scale for a wide range of the 
control parameters. For example, it has been proven by Engle that
the ARCH(1) process has finite variance for $\alpha_1<1$, and finite 
fourth moment for $3 \alpha_1^2<1$ \cite{Engle82}.  

We simulate several ARCH(1) process to investigate the dynamics of
the unconditional probability density function $P(Z_{n\Delta t})$ at different time horizons $n\Delta t$ ($Z_{n\Delta t} \equiv \sum_{i=1}^n Z_{t-i}$). For each simulation
we also study the scaling properties of the ``probability of return
to the origin" $P(Z_{n\Delta t}=0$) as a function of $n\Delta t$. We select the values
of the control parameters to investigate ARCH(1) processes which are characterized by the same unconditional variance observed in our empirical investigation
of the S\&P500 dynamics (namely $\sigma^2=2.57 \cdot 10^{-3}$) and by
different values of the kurtosis $\kappa$ of the $\{ Z \}$ process.
For an ARCH(1) process the unconditional variance is given by \cite{Engle82}
\begin{equation}
\sigma^2=\frac{\alpha_0}{1-\alpha_1}
\end{equation} 
while the kurtosis is \cite{Engle82}
\begin{equation}
\kappa=\frac{3(1-\alpha_1^2)}{1-3\alpha_1^2}
\end{equation} 

We focus our attention on three cases: 

(i) $\alpha_0=0.00231$ and
$\alpha_1=0.1$ . In this case $\sigma^2=2.57 \cdot 10^{-3}$ and 
$\kappa=3.06$. The value of $\kappa$ is very close to the one 
expected for a Gaussian stochastic process ($\kappa=3$); 

(ii) $\alpha_0=0.00112$ and $\alpha_1=0.564$. With these values of
the control parameters the variance and the kurtosis are $\sigma^2=2.57 \cdot 10^{-3}$ and $\kappa=43$. This value of $\kappa$ is approximately the same value observed in the empirical analysis of the S\&P500 changes
for time intervals $\Delta t=1$ minute; 

(iii) $\alpha_0=0.00109$ and
$\alpha_1=0.575$ . Values of the control parameters implying the 
same variance as above but a very high value for the kurtosis 
($\kappa=247$).

By varying the values of the control parameters, it is of course
possible to make the shape of the PDF $P(Z)$ more leptokurtic than
a Gaussian distribution. The presence of a given degree of leptokurtosis
does not imply directly scaling properties of $P(Z_{n\Delta t})$ PDFs 
strongly different from the Gaussian scaling. By studying the 
``probability of return to the origin", we find that
an approximate scaling behavior is present in ARCH(1) stochastic
processes for short times
($n\Delta t \le 100$). We find that the values of the scaling exponent best describing the above-cited time evolution of $P(Z_{n\Delta t}=0)$ are 2.02, 1.93 and 1.85 respectively. These values are very close 
to the scaling exponent 2 observed for a Gaussian stochastic process. 
Hence an ARCH(1) process is not able
to describe the scaling properties empirically observed in the 
stochastic dynamics of the S\&P500 for $\Delta t < 1000$ minutes
(where the scaling exponent is 1.4).

\section{GARCH process}

ARCH(1) model is the simplest autoregressive model. In the following,
we will consider a less simple autoregressive model, the GARCH(1,1)
model. The GARCH(1,1) model is widely studied in the economic literature
\cite{Bollerslev92}.
In 1986 generalized ARCH or GARCH(p,q) models were proposed 
\cite{Bollerslev86}. These models are more flexible than ARCH models
in the lag structure. They are defined by the relation
\begin{equation}
\sigma^2_t=\alpha_0+\alpha_1 Z_{t-1}^2+ ....... + \alpha_p Z_{t-p}^2
+\beta_1 \sigma_{t-1}^2+ ....... + \beta_q \sigma_{t-q}^2
\end{equation}
where the constants $\alpha_0, .... , \alpha_p, \beta_1, .... , \beta_q$
are the control parameters of the GARCH stochastic process. The simplest
GARCH process, the GARCH(1,1), is often studied in the modeling of
prices of financial assets. GARCH(1,1) processes are unconditional
stationary with finite variance and fourth moment if $1-\alpha_1-\beta_1>0$
and $1-\beta_1^2-2 \alpha_1 \beta_1 - 3 \alpha_1^2 > 0$ respectively.
Empirical analyses of stock market price data have shown (see for 
example  \cite{Akg89}) that a good choice of the parameter $\beta_1$
is $\beta_1=0.9$ . Accordingly,  we set $\beta_1=0.9$ in our simulations 
and we set the remaining control parameters $\alpha_0$ and $\alpha_1$ to the values $\alpha_0=2.3 ~~ 10^{-5}$ and $\alpha_1=0.09105$. With this choice of control parameters, the unconditional variance \cite{Baillie92} 
\begin{equation}
\sigma^2=\frac{\alpha_0}{1-\alpha_1-\beta_1}
\end{equation} 
of the process $\{Z\}$ is approximately equal to the value observed 
in the S\&P500 data. The kurtosis \cite{Baillie92} 
\begin{equation}
\kappa=\frac{6\alpha_1^2}{1-\beta_1^2-2 \alpha_1 \beta_1-3\alpha_1^2}+3
\end{equation} 
also assumes the value measured in the empirical analysis of the S\&P500
data ($\kappa=43$). The simulated GARCH(1,1) process has an unconditional
PDF $P(Z_{\Delta t})$ which mimics very well the one observed in the S\&P500 data 
with a time interval $\Delta t=1$ minute \cite{Mantegna95}. We also study the probability
of return to the origin to determine if scaling is observed for the 
PDFs of this process. We observe a scaling behavior for a wide range of time ($n\Delta t < 10,000$). The measured scaling exponent is 1.88, a value close to the Gaussian scaling exponent and
rather different from the scaling exponent found in empirical data
($\alpha=1.4$).

In summary, the GARCH(1,1) process fails to properly describe the scaling properties
of the S\&P500 index detected for $\Delta t < 1000$ minutes. However, GARCH(1,1) is able to give an accurate description of the 
$\Delta t=1$ minute PDF using as control paremeters $\beta_1=0.9$ and
obtaining the values of $\alpha_0$ and $\alpha_1$ from the 
values of $\sigma^2$ and $\kappa$ measured from the empirical data.
  
\section{Discussion}

Our study shows that the problem of the {\it complete} stochastic
characterization of index (or price) dynamics in a financial market
is an open question. For example, both models considered here have strengths and 
limitations, and modifications of them are needed to reach a more 
satisfactory agreement with the results of empirical analyses. The problem
of stochastic modeling of price dynamics comprises fundamental and applied 
aspects. The fundamental aspects are related to the theoretical modeling
of a nonlinear {\it complex system} evolving without known conservation
laws in the presence of quenched and external noise. The applied aspects
are related to the role that the exact shape of the PDF of stock returns
and the time evolution of the variance of stock returns (volatility
in the economic literature \cite{Schwert89}), plays in the pricing of
derivative financial products \cite{Hull97}. An extremely important 
activity performed everyday in financial
markets.

We think that the mixing of empirical analyses, modeling, simulations
and comparison between empirical data and simulations constitutes
a scientific procedure that will allow us to eventually find the most
accurate and ``parsimonious" stochastic model describing index
(or price) dynamics.  

\acknowledgements

We thank INFM, MURST and NSF for financial support.

\end{document}